\documentclass[pdflatex,sn-mathphys-num]{sn-jnl}

\usepackage{graphicx}%
\usepackage{multirow}%
\usepackage{amsmath,amssymb,amsfonts}%
\usepackage{amsthm}%
\usepackage{mathrsfs}%
\usepackage{xcolor}%
\usepackage{textcomp}%
\usepackage{manyfoot}%
\usepackage{booktabs}%
\usepackage{listings}%

\begin{document}

\title[SMARTHEP]{SMARTHEP: training PhD students in real-time analysis at the LHC and in industry}


\author[1]{\fnm{Johannes} \sur{Albrecht}}
\author[2,3]{\fnm{Laura} \sur{Boggia}}
\author[4]{\fnm{Leon} \sur{Bozianu}}
\author[5]{\fnm{Andrew} \sur{Carey}}
\author[6]{\fnm{Carlos} \sur{Cocha}}
\author*[5]{\fnm{Caterina} \sur{Doglioni}}\email{caterina.doglioni@cern.ch}
\author[1]{\fnm{James Andrew} \sur{Gooding}}
\author[7]{\fnm{Joachim} \sur{Hansen}}
\author[8]{\fnm{Patin} \sur{Inkaew}}
\author[7]{\fnm{Kaare} \sur{Iversen}}
\author[5]{\fnm{Pratik} \sur{Jawahar}}
\author[8,9]{\fnm{Henning} \sur{Kirschenmann}}
\author[10,11]{\fnm{Daniel} \sur{Magdalinski}}
\author*[7]{\fnm{Alice} \sur{Ohlson}}\email{alice.ohlson@cern.ch}
\author[1]{\fnm{Micol} \sur{Olocco}}
\author[12,13]{\fnm{Henrique} \sur{Pi\~{n}eiro Monteagudo}}
\author[4]{\fnm{Steven} \sur{Schramm}}
\author[14]{\fnm{Mike} \sur{Sokoloff}}
\author[7]{\fnm{Alexandros} \sur{Sopasakis}}
\author[12]{\fnm{Leonardo} \sur{Taccari}}
\author[7]{\fnm{Sten} \sur{{\AA}strand}}

\affil[1]{\orgname{Technical University of Dortmund}, \orgaddress{\city{Dortmund}, \country{Germany}}}
\affil[2]{\orgname{IBM France}, \orgaddress{\city{Paris}, \country{France}}}
\affil[3]{\orgname{Sorbonne University}, \orgaddress{\city{Paris}, \country{France}}}
\affil[4]{\orgname{University of Geneva}, \orgaddress{\city{Geneva}, \country{Switzerland}}}
\affil[5]{\orgname{University of Manchester}, \orgaddress{\city{Manchester}, \country{United Kingdom}}}
\affil[6]{\orgname{Heidelberg University}, \orgaddress{\city{Heidelberg}, \country{Germany}}}
\affil[7]{\orgname{Lund University}, \orgaddress{\city{Lund}, \country{Sweden}}}
\affil[8]{\orgname{University of Helsinki}, \orgaddress{\city{Helsinki}, \country{Finland}}}
\affil[9]{\orgname{Lappeenranta-Lahti University of Technology LUT}, \orgaddress{\city{Lahti}, \country{Finland}}}
\affil[10]{\orgname{NIKHEF}, \orgaddress{\city{Amsterdam}, \country{Netherlands}}}
\affil[11]{\orgname{VU Amsterdam}, \orgaddress{\city{Amsterdam}, \country{Netherlands}}}
\affil[12]{\orgname{Verizon Connect}, \orgaddress{\city{Florence}, \country{Italy}}}
\affil[13]{\orgname{University of Bologna}, \orgaddress{\city{Bologna}, \country{Italy}}}
\affil[14]{\orgname{University of Cincinnati}, \orgaddress{\city{Cincinnati}, \state{OH}, \country{USA}}}


\abstract{
In this Editorial, we describe the SMARTHEP Innovative Training Network funded via the Marie Sk{\l}odowska-Curie Actions between 2021 and 2025. SMARTHEP trained 12 PhD students to advance machine learning and real-time analysis in high-energy physics experiments and industrial applications. We present the perspective of students, supervisors, and external observers of the network, concerning the work done within the network, the added value compared to ``typical'' PhD positions, and the emerging themes and directions from our experiences in the past four years. 
}


\maketitle

\section{Real-time analysis in high-energy physics and industry}

The volume of data currently produced and available to both scientific research and industrial sectors is increasing at an unprecedented rate~\cite{10.3389/fdata.2023.1271639}. However, this exponential growth in data collection is not always matched by comparable advancements in storage, utilisation, or analysis capabilities. 
When the capacity to store information is limited, data must inevitably be discarded or simply not recorded. 
This issue is particularly acute in High-Energy Physics (HEP), where experiments at the Large Hadron Collider (LHC) generate hundreds of gigabytes of data every second.
It is financially and technically unviable to retain this volume of information within current storage resource limitations, necessitating the development of novel techniques to:
\begin{itemize}
\item Enable the LHC experiments' \emph{trigger} systems~\cite{Gooding:2024wpi} to make rapid, high-fidelity decisions on which particle collision events should be written to short- or long-term storage for future analysis, or discarded.
\item Shift part or all of the data acquisition workflow from a ``record first, analyse later'' model to a Real-Time Analysis (RTA) approach~\cite{CMS:2024zhe,Aaij:2019uij,ATLAS:2025okg,Buncic:2015ari}, where substantial analysis is performed as close to the detector as possible.
This makes it possible to reduce the volume of data saved per collision for further analysis, instead of significantly reducing the number of stored events, as is the case for traditional triggers. 
\end{itemize}
Commercial applications currently face a parallel challenge: the complexity of available data grows rapidly, while the resources and time available to make decisions based on that information do not scale accordingly. 
Hence, data-taking and data analysis processes in both research and industry must become significantly more efficient to use this wealth of information in a cost- and resource-effective manner. 
Meeting this challenge requires the training of a new generation of researchers with cross-disciplinary expertise, bridging the gap between problems related to fundamental research and the cutting-edge algorithms used in modern industry. 
Training Early Stage Researchers (ESRs) to navigate this intersection has been the primary objective of the SMARTHEP network.

\section{Goals and structure of SMARTHEP}

The Marie Sk{\l}odowska-Curie Innovative Training Network (MSCA ITN) SMARTHEP -- \textit{Synergies between Machine leArning, Real-Time analysis and Hybrid architectures for efficient Event Processing and decision making} (Call: H2020-MSCA-ITN-2020, GA 956086) -- was established to train a cohort of ESRs proficient in the synergistic application of Machine Learning (ML) to RTA across both scientific and industrial domains. 
The SMARTHEP ESRs contributed directly to the LHC physics programme by designing and implementing the trigger systems and calibration algorithms used in the ATLAS, CMS, LHCb, and ALICE experiments.
Concurrently, they leveraged their experience to deliver concrete commercial software and algorithms in fields that require rapid decision-making, such as fleet safety, manufacturing, and finance.
Through a structured system of training and secondments (see Sec.~\ref{sec:secondments}), SMARTHEP enabled exchanges between twelve European academic institutions and a diverse range of industrial partners, ranging from specialised start-ups to multinational corporations.
This collaborative framework ensured that advancements in RTA from the high-energy physics environment could feed into real-world applications and vice versa, thereby satisfying the growing demand for cross-cutting expertise in big data analysis.

The SMARTHEP network supported twelve ESRs across eight European nations, all enrolled in doctoral programs. 
The ESRs conducted their PhD either within the four major LHC collaborations or at industrial partners such as IBM and Verizon Connect. 
The students received dual supervision by experts from both academia and the private sector, reinforced by the annual network meetings and training events that took place in Manchester, CERN, Lund, Milan, and Dortmund.
These events, and the regular monthly management meetings with participation of both supervisors and ESRs, established a cross-experimental framework that provided value exceeding standard institutional doctoral positions. 
As noted by one ESR:
\begin{quote} \textit{``SMARTHEP provides opportunities that go well beyond a regular PhD studentship, as it links the four LHC experiments with industry partners.
We gained advanced technical skills, industry exposure, and benefited from dedicated financial resources that gave us greater independence to pursue valuable learning and networking opportunities.
A unique strength was the cross-experimental community, where students regularly contributed to solving common technical challenges.
This combination of research excellence, industry engagement, and enriched resources represents clear added value compared to a traditional PhD.''} \end{quote}
This sentiment was corroborated by the External Advisory Board, composed of world renowned experts in RTA whose task was to monitor the network's efficacy and progress. 
One Board member observed that the students \textit{``developed expertise in software for real-time applications in the private sector that complemented their experiences in the HEP world,''} demonstrating the utility of building communities around shared technical interests.
From the perspective of the Principal Investigators (PIs) in the LHC collaborations, the network offered two primary advantages. 
First, it resulted in significant contributions to the LHC trigger systems; the funding model encouraged ESRs to dedicate a significant fraction of their time to technical RTA topics, a commitment often infeasible in traditional funding schemes that favour physics analysis. 
Second, it fostered a cohesive community of PIs and students working on RTA topics, facilitating high-level technical exchange while adhering to the data sharing protocols of individual collaborations.

\section{Student and supervision experience}
\label{sec:secondments}


Integral to the SMARTHEP training mandate was the secondment mechanism, where students spent a period of several months under the supervision of a PI at another node of the network. 
This resulted in a structured opportunity for ESRs to acquire dual-sector competencies and career development, as well as an expanded professional network and potential hiring pipeline for the industrial partners.
The experience of industrial partner Ximantis AB, which hosted two ESRs, illustrates the nuance of these exchanges: while the collaboration demanded a significant time investment relative to immediate commercial gain, it yielded tangible scientific outputs, including joint articles~\cite{giasemis2025learningtrafficanomaliesgenerative,Ximantis2}, and fostered network cohesion. 
For the seconded students, these placements provided exposure to Graph Neural Networks (GNNs)  \cite{Sopasakis2, Sopasakis1} within an agile industry environment, a context distinct from large-scale scientific collaborations.
Conversely, the ``reverse'' secondment of an industry-based ESR from Verizon Connect into academia highlighted deep synergies in real-time constraints; specifically, the necessity for model optimisation techniques such as quantisation, pruning, and knowledge distillation to maximise throughput on hardware. 
While domain-specific barriers remain, such as the steep learning curve associated with HEP-specific jargon and frameworks, the use of standard ML libraries (e.g., PyTorch) lowers these hurdles.
The demonstrable success of this mobility framework has had a lasting institutional impact, incentivising supervisors to establish analogous secondment programmes at their home faculties, thereby validating the SMARTHEP and MSCA ITN model as a standard for doctoral training.


The success of the SMARTHEP training model is also demonstrated by the professional growth of the ESRs within the LHC collaborations. 
For instance, at the LHCb experiment, one ESR progressed from developing specific software tools to a leadership position managing one of the collaboration's most data-intensive channels; they are now responsible for optimising resources and ensuring the quality of data collection.
A similar trajectory occurred within the CMS experiment: One of the ESRs, with no prior experience in HEP, started by contributing to the monitoring of the RTA stream; by the end of their PhD they were the coordinator fully responsible for its operation.
These examples demonstrate how technical training networks effectively contribute to developing students into the leaders needed to maintain the critical real-time infrastructure of particle physics experiments.


The SMARTHEP network extended its reach by including several ``affiliated students'': doctoral candidates enrolled at the network’s partner institutes but supported by external funding sources. 
These students were fully integrated into the network's ecosystem, participating in the network’s training sessions and annual meetings alongside the core cohort. 
This inclusive structure allowed them to discuss their projects with a group of expert PIs, industry partners and peers not typically available at a single university, and enhanced their research outcomes beyond what would have been possible without the network.
This was beneficial for the SMARTHEP environment itself as the ESR and affiliated student cohort functioned as a team, e.g., when participating in hackathons or in projects at summer schools, building lasting collaborations that extend beyond the lifetime of the project. 

\section{Benefits for experiments and industry involved in the network}

In terms of research achievements for the LHC experiments, ESRs in the SMARTHEP network focused on the optimisation of the experiments' trigger systems and the use of RTA techniques in physics analysis.
Examples of technical outcomes include the benchmarking of FPGAs against GPUs for ML-based track reconstruction in LHCb to evaluate throughput in hybrid architectures, the application of GNNs for track finding within the LHCb Vertex Locator, anomaly detection frameworks utilising synthetic data generation for calorimeter time series, and RTA strategies to facilitate the retention of specific data subsets that would otherwise have exceeded standard storage constraints. 
The ESRs also wrote and published two whitepapers, one summarising the trigger systems of the four major experiments, and another one on the use of ML in RTA systems~\cite{Albrecht:2024eio,Boggia:2025kbs}.
ESRs also played leading roles in the successful coordination of these whitepapers, developing soft skills in project management.

In terms of industrial and commercial applications, the network advanced RTA methodologies in autonomous driving, traffic modelling and finance. 
Example outcomes include the methods developed by an ESR at Verizon Connect, who created an object detection algorithm capable of identifying lane changes and overtaking manoeuvres from motion profiles~\cite{pineiro2023object}, and studied techniques to generate bird's-eye-view representations of driving scenes with few or no labelled data~\cite{pineiro25rendbev}.
Many of these outcomes involved the transfer of physics-based ML and analysis techniques to decision-making problems in industry and society. 
The network's collaborative framework fostered collaborations between research and industry, for example where two ESRs used anomaly detection in both the Data Quality Monitoring system of one of the experiments and in financial transactions. 
By converging on a shared dataset, the ESRs developed two distinct algorithms, grounded in financial fraud detection techniques and information theory principles, that can be benchmarked against baseline references. 

More generally, the network benefitted from regular communication between ESRs from industrial and academic backgrounds.
The writing of the aforementioned whitepaper on ML in RTA involved contributions from and discussions between both communities, which helped both sides to understand how to better communicate with people from different domains, and where they may be making assumptions in their statements.
This effort also helped the ESRs to better understand how the terminology they were using could be interpreted in different ways, and led to extensive discussions on defining what it meant for an analysis to be performed in a real-time environment.
These interactions helped to highlight that while the LHC and industrial applications may have different associated timescales or types of input data, they do have common challenges, and thus a lot can be learned by working together.

\section{Outreach and communication}

The network had an active outreach component, including local events and public talks as part of the yearly meetings. 
Notably, as part of the SMARTHEP outreach hackathon that took place in June 2024, the SMARTHEP ESRs contributed to the INFN-led outreach project HEPscape!~\cite{HEPSCAPE}, a HEP-themed escape room kit for the general public to learn about high-energy physics and accelerators.
The international nature of the network was instrumental for translating the original Italian videos and material into several of the network languages (English, Spanish, German, Thai) and preparing for the delivery of the escape room in the ESR's native countries~\cite{soffi_2024_14575673}. 

Co-organisation of events, conferences and schools was possible through a dedicated budget line item in SMARTHEP. This made it possible to spread the network’s message beyond standard HEP conferences, and to sponsor events aligning with the network’s goals.
Among others, SMARTHEP sponsored the following conferences: 
\begin{itemize}
    \item \textit{EuCAIFCon (2024, 2025)}: The sponsorship of the first two editions of the European Coalition for AI in the Fundamental Sciences (\href{EuCAIFCon}{https://eucaif.org}) strengthened the synergies between the network and the coalition. 
    Since AI and ML constitute one of the key methodologies of SMARTHEP, this connection ensured visibility for the network and enabled the ESRs to serve on the Local Organising Committee and present their technical deliverables to a broad, cross-disciplinary audience.
    \item \textit{Workshop on workflow languages for HEP analysis (2025)}: The sponsorship of this one-off workshop organised by \href{FAIROS-HEP}{https://fairos-hep.org} and the \href{HEP Software Foundation}{https://hepsoftwarefoundation.org} enabled further integration of the network into the broader scientific computing community.
    This workshop brought together expertise from the LHC experiments and an exchange of knowledge on data analysis techniques entirely aligned with the goals of the network.
    \item \textit{SSI Collaborations Workshop (2025)}: The sponsorship of the annual workshop organised by the UK’s \href{Software Sustainability Institute}{https://www.software.ac.uk} allowed the network to present its objectives to the wider Research Software Engineering (RSE) community.
    This engagement established critical connections regarding software training and environmental sustainability that have become an integral part of the network and of its future. 
\end{itemize}
In 2024, SMARTHEP also co-organised a school together with the Next Generation Triggers (NextGen) project at CERN. NextGen is a collaboration between CERN and the ATLAS/CMS experiments with the key objective to get more physics information out of LHC data, with a strong focus on the experiments’ trigger systems.
The \textit{SMARTHEP Edge Machine Learning School} taught both SMARTHEP and NextGen Fellows cohorts how to deploy ML on edge hardware such as FPGAs and GPUs to meet the latency constraints of future trigger systems.  

In order to communicate and strengthen links with the broader industrial sector, one of the ESRs in the SMARTHEP network led the organisation of the \textit{SMARTHEP meets Industry @ CERN} event to demonstrate that skills developed in high-energy physics -- from real-time data processing to ML on heterogeneous architectures -- are also relevant to broader society. 
In collaboration with Confindustria Piemonte, Fondazione Piemonte Innova, the Digital Innovation Hub, and Enterprise Europe Network, the initiative attracted twenty-five companies from Italy and France, representing sectors as diverse as Industry 4.0, finance, and the automotive industry. 
Prior to the event, the network distributed detailed catalogues of ESR activities and projects to these companies, which enabled targeted one-on-one networking sessions focused on technology transfer. A dedicated brochure was also prepared for the invited industries and SMARTHEP’s own industrial partners, ensuring a two-way exchange of information. 
Hosting the meeting at CERN highlighted the connection between fundamental research and commercial innovation, a theme reinforced in a presentation by a representative of CERN’s Knowledge Transfer department. 
This event format provided a practical forum for direct dialogue between researchers and industry, which can be adopted for future industrial/academic network collaborations. 

\section{Emerging themes}

Throughout the lifetime of the network, the PIs and ESRs had the chance to reflect on a number of topics beyond the technical tasks at hand. We summarise a few of these reflections in the following sections. 

\subsection{Ethics considerations in ML}

While there may be no ethical considerations when analysing data from particle collisions at the LHC, it has been essential for ESRs to consider potential ethical implications when handling personal data in their industrial PhD projects and secondments. 
SMARTHEP benefitted from a dedicated Ethics Advisor to provide guidance and training to the network, particularly focused on biases that can arise when ML techniques are transferred from HEP to the analysis of datasets containing information about human actions (e.g., real-time monitoring of traffic patterns or financial fraud detection).  
When a ML model is trained or makes decisions based on datasets including personal data (even anonymised), it is vital to establish precautions to eliminate or mitigate potential algorithmic biases, and to put in place procedures for the model to be able to justify the results it provides in specific situations. 
As the SMARTHEP ESRs embark on their careers in a world in which increasingly more ML/AI models make use of our personal data, it is crucial to establish a habit and culture of considering and mitigating ethical risks from the very beginning.  

\subsection{Software sustainability in HEP collaboration}

Sustaining the complex software that underpins modern scientific experiments requires more than just technical expertise; it demands stable, long-term funding and a collaborative, networked approach~\cite{HEPSoftwareFoundation:2025irs}.
Much of the essential work in software development -- such as maintenance, documentation, testing, and upgrades of infrastructure -- is critical for the reliability and longevity of experimental research, but is not always supported by funders and academic employers who prioritise scientific novelty and output in the form of high-profile publications. Stable funding streams and career paths are vital to supporting this ongoing work, enabling the community to maintain, evolve, and innovate the software that is needed for scientific discovery.

\subsection{The emergence of Large Language Models}

The rapid emergence of Large Language Models (LLMs) while SMARTHEP was ongoing prompted many discussions on their utility in both science and industry. 
In the commercial sector, a secondment of one of the ESRs at IBM France Lab applied physics methodologies -- inspired by simulations in particle physics -- to financial sequences; this project explored LLMs to generate synthetic transaction logs, creating essential training data for fraud detection systems. 
Simultaneously, within the high-energy physics domain, students in the network applied these generative tools to software development, focusing on the use of LLMs to optimise complex real-time trigger code; this was done by ESRs who were already experts in both the programming language and the trigger chains. 
Looking ahead, the concept of Large Physics Models (LPMs) is gaining momentum, envisioning models specifically tailored to the needs of physics research, which could offer a unified framework for tackling the unique challenges of the field~\cite{Barman:2025wfb} together with broader infrastructure needs~\cite{Caron:2025rir}. 

\subsection{Next steps towards RTA on the edge}

A research direction stemming from SMARTHEP aims to transition real-time AI from traditional computing clusters directly to the ``edge'' of enhanced experimental detectors, i.e., as close to the detector as possible. 
As data generation rates and complexity in fundamental physics continue to grow, it is worth investigating how ML algorithms can be embedded directly into sensor electronics, and how the additional features provided by new detector technologies can further improve insight from data. 
Here, training is needed on the co-design of algorithms and hardware, enabling data reduction decisions to be made even earlier in the processing pipeline. 
Many of the SMARTHEP PIs are now involved in the COST Action CA42153 EPIGRAPHY (\url{https://www.cost.eu/actions/CA24153/}), focused on meeting the challenges presented by edge algorithms, which include strong constraints on latency and the available computing resources. 

\section{Conclusions}

The experience of the SMARTHEP network demonstrated that the gap between fundamental research and industrial application can be bridged by a shared focus on real-time analysis and machine learning. 
By placing Early Stage Researchers at the intersection of these fields, the SMARTHEP network provided training enabling them to manage the data challenges of the High-Luminosity LHC while also driving innovation in sectors like autonomous systems and finance. 
The results -- ranging from new trigger algorithms at ATLAS, CMS, and LHCb, to improved commercial fraud detection systems -- prove that cross-sector training can generate tangible technical benefits for all participants.
Moreover, the partnerships established with the EuCAIF coalition, the NextGen project and the EPIGRAPHY COST Action ensure that these training methods will continue beyond the current funding cycle. 

\backmatter

\bmhead{Acknowledgements}

This work is part of the SMARTHEP network, funded by the European Union’s Horizon 2020 research and innovation programme under Grant Agreement No. 956086.

\section*{Declarations}

\subsection*{Disclosure of Delegation to Generative AI}

The authors declare the use of generative AI in the research and writing process. According to the GAIDeT taxonomy \cite{Suchikova08082025}, the following tasks were delegated to the generative AI tool Gemini 3 Pro under full human supervision:

\begin{itemize}
\item Proofreading and editing
\item Summarising text
\item Adapting and adjusting tone
\end{itemize}

We have used a specialised Gemini 3 model trained on our writing (papers, past editorials and a previous version of the SMARTHEP proposal) to unify the style of texts composed and contributed by different authors, and to summarise parts of original contributed text. The model was explicitly instructed to serve as a style guide rather than as a text generator drawing from the sources. All authors, as well as all network participants, carefully reviewed and edited the resulting text to ensure that it aligned with their original work.

Responsibility for the final manuscript lies entirely with the authors, generative AI tools are not listed as authors and do not bear responsibility for the final outcomes.


\subsection*{Competing interests}
The authors have no competing interests to declare that are relevant to the content of this article.

\bibliography{sn-bibliography}

\end{document}